\documentclass[fp,twocolumn]{jpsj3}
\usepackage{amssymb}
\usepackage{txfonts}

\inst{$^1$Key Laboratory for Microstructural Material Physics of
Hebei Province, School of Science, Yanshan University, Qinhuangdao 066004, China \\
$^2$Department of Physics and Astronomy, University of Pittsburgh,
Pittsburgh,Pennsylvania 15260, USA \\
$^3$Department of Physics, Tangshan Normal University, Tangshan 063000, China\\
$^4$Department of Physics, Xinzhou Teachers University, Xinzhou 034000, China}
\abst{We investigate the topological defects and spin structures of
binary Bose-Einstein condensates (BECs) with Dresselhaus spin-orbit
coupling (D-SOC) in a rotating anharmonic trap. Our results show
that for initially mixed BECs without SOC the increasing rotation
frequency can lead to the structural phase transition of the system.
In the presence of isotropic D-SOC, the system sustains vortex pair,
Anderson--Toulouse coreless vortices, circular vortex sheets, and
combined vortex structures. In particular, when the rotation
frequency is fixed above the radial trapping frequency the strong
D-SOC results in a peculiar topological structure which is comprised
of multi-layer visible vortex necklaces, hidden vortex necklaces and
a hidden giant vortex. In addition, the system exhibits rich spin
textures including basic skyrmion, meron cluster, skyrmion string
and various skyrmion lattices. The skyrmions will be destroyed in
the limit of large D-SOC or rotation frequency. Furthermore, the
effects of anisotropic D-SOC and Rashba-Dresselhaus SOC on the
topological structures of the system are discussed.}

\begin{document}

\title{Topological defects of spin-orbit coupled Bose-Einstein condensates
in a rotating anharmonic trap}
\author{Chunxiao Shi$^1$, Linghua Wen$^1$$^,$$^2$\thanks{%
linghuawen@ysu.edu.cn}, Qingbo Wang$^1$$^,$$^3$, Hui Yang$^1$$^,$$^4$, and
Huan Wang$^1$}
\maketitle

\section{Introduction}

One of the most exciting recent developments in cold atom physics has been
the production of spin-orbit-coupled quantum gases \cite{Lin,Cheuk,Meng,ZWu}%
. The spin-orbit coupling (SOC) between a quantum particle's spin and its
momentum in solid materials is essentially an intrinsic property of the
materials and can not be controlled due to the unavoidable impurities and
disorder \cite{Xiao,LiX}. By contrast, the SOC in neutral atomic gases can
be controlled effectively through a series of experimental parameters \cite%
{Lin,Cheuk,Meng,ZWu,HuangL,Kolkowitz,LiJ}, which means that the ultra-cold
atomic gases with SOC provide an ideal platform to explore the novel quantum
states and relevant dynamics \cite%
{WangC,Sinha,Hu,Ramachandhran,Kawakami,ZhangY,Zhu,XuY,White,Qu,Sakaguchi1,Sakaguchi2,Stringari,Khamehchi,Kartashov,Sakamoto}%
. Typically, the nonrotating spin-orbit coupled Bose-Einstein condensates
(BECs) in a harmonic trap support plane-wave phase \cite{ZhangY}, stripe
phase \cite{WangC,ZhangY,LiJ}, half-quantum vortex \cite%
{Sinha,Hu,Ramachandhran}, bright soliton \cite{Sakaguchi2,XuY2,Gautam}, dark
soliton\cite{Achilleos,XuY3}, gap soliton \cite{ZhangY2}, and topological
superfluid phase \cite{ZWu}. In particular, the combination effect of SOC
and rotation on BECs has been shown to be able to create various topological
defects. These intriguing results enrich the phase diagram and physics of
condensate system. To the best of our knowledge, however, most literature
concerning on the rotating spin-orbit coupled BECs focuses on the harmonic
trap \cite{XuX,Zhou,Radic,Aftalion,Fetter,XuZ}. In the case of rotating
harmonic trap, the rotation angular frequency $\Omega $ is generally limited
to a relatively small value far below the radial trapping frequency $\omega
_{\bot }$. In the limit $\Omega /\omega _{\bot }\rightarrow 1$, a rotating
BEC is expected to undergo complicated quantum phase transition from
superfluid to various highly correlated ground states (nonsuperfluid
states). Indeed, this situation is singular because the total angular
momentum and the Thomas-Fermi (TF) radius of the BEC both diverge.

In actual experiments, the trap usually is not purely harmonic. With this
concern, we study the ground-state structures of spin-orbit coupled BECs in
a rotating anharmonic trap (a rotating harmonic trap with a quartic
distortion) \cite{Fetter2}. Such an anharmonic trap can confine the BECs
even for $\Omega /\omega _{\bot }>1$, and therefore allowing a more
controlled investigation of possible new states which are not generally the
same as those expected in a harmonic trap. For the spin-orbit interaction,
here we mainly consider the Dresselhaus SOC (D-SOC) and Rashba-Dresselhaus
SOC (RD-SOC) which can be realized under current experimental conditions
\cite{Lin,Cheuk,Meng,ZWu,HuangL,Kolkowitz,LiJ,Qu,Dresselhaus,Goldman}. For
initially mixed two-component BECs without SOC, we show that structural
change of vortex patterns can be achieved by regulating the rotation
frequency. In the presence of SOC, the system can exhibit rich vortex
defects and various kinds of skyrmion structures \cite{Skyrme,Zhai,LiuC1}.
Particularly, for the case of large D-SOC strength or rotation frequency,
the skyrmion structures will be destroyed due to the overlap between the two
components. In addition, the combined effects of SOC, rotation, interatomic
interactions and anharmonic trap are revealed and discussed. Furthermore, we
find that there are evident differences for the vortex structures and spin
textures between the rotating BECs with D-SOC and those with RD-SOC.

The paper is organized as follows. In section 2, the theoretical model is
introduced and the coupled dynamic equations are given. The topological
structures and relevant spin textures of the system are presented and
analyzed in section 3. Our findings are summarized in the last section.

\section{Model}

By assuming tight confinement in the $z$ direction, we consider a
quasi-two-dimensional system of rotating two-component BECs with D-SOC which
are confined in an anharmonic trap (i.e., a harmonic plus quartic trap). The
energy functional of the system is given by
\begin{eqnarray}
E &=&\int \sum\limits_{j=1,2}\left\{ -\frac{\hbar ^{2}}{2m}\left\vert \nabla
\psi _{j}\right\vert ^{2}+V_{tr}\mathbf{(}r\mathbf{)}\left\vert \psi
_{j}\right\vert ^{2}+\frac{g_{j}}{2}\left\vert \psi _{j}\right\vert
^{4}\right.  \nonumber \\
&&\left. +\hbar \psi _{j}^{\ast }\left[ -i\lambda _{x}\frac{\partial \psi
_{3-j}}{\partial x}+\left( -1\right) ^{j}\lambda _{y}\frac{\partial \psi
_{3-j}}{\partial y}\right] -\Omega \psi _{j}^{\ast }L_{z}\psi _{j}\right\}
\mathrm{{d}^{2}r}  \nonumber \\
&&+\int g_{12}\left\vert \psi _{1}\right\vert ^{2}\left\vert \psi
_{2}\right\vert ^{2}\mathrm{{d}^{2}r,}  \label{1-1}
\end{eqnarray}%
where $\psi _{1,2}$ denote the two-component (i.e., spin-up and spin-down)
wave functions, and they are normalized as $\int [|\psi _{1}|^{2}+|\psi
_{2}|^{2}]\mathrm{{d}x{d}y=\emph{N}}$ with $N$ being the total particle
number. $g_{j}=4\pi a_{j}\hbar ^{2}/m$ $\left( j=1,2\right) $ and $%
g_{12}=2\pi a_{12}\hbar ^{2}/m$ denote the intra- and intercomponent
interaction strengths that are characterized by the corresponding $s$-wave
scattering lengths $a_{j}$ and $a_{12}$ between intra- and interspecies
atoms and the atomic mass $m$. $\Omega $ is the rotation frequency along the
$z$ direction, and $L_{z}=i\hbar (y\partial _{x}-x\partial _{y})$ denotes
the $z$ component of the angular-momentum operator. The D-SOC is written as$%
\ v_{D}=i\hbar (\lambda _{x}\sigma _{y}\partial _{x}+\lambda _{y}\sigma
_{x}\partial _{y})$ \cite{Lin}, where $\sigma _{x,y}\ $are the Pauli
matrices and $\lambda _{x}$ and $\lambda _{y}$ are the SOC strengths in the $%
x$ and $y$ directions. The anharmonic trap $V_{tr}(r)$ \cite{Fetter2} is
described by%
\begin{equation}
V_{tr}\left( r\right) =\frac{1}{2}m\omega _{\perp }^{2}\left( r^{2}+\mu
\frac{r^{4}}{a_{0}^{2}}\right) =\frac{1}{2}\hbar \omega _{\perp }\left(
\frac{r^{2}}{a_{0}^{2}}+\mu \frac{r^{4}}{a_{0}^{4}}\right) ,  \label{1-2}
\end{equation}%
with $\omega _{\perp }$ being the radial trap frequency and $a_{0}=\sqrt{%
\hbar /m\omega _{\perp }}$ being the harmonic-oscillator length. Here $r=%
\sqrt{x^{2}+y^{2}}$ is the radial coordinate in two-dimensions, and $\mu $
is a dimensionless constant that characterizes the anharmonicity of the
trap. For the sake of numerical calculation, it is convenient to introduce
the following notations $\widetilde{r}=r/a_{0}$, $\widetilde{t}=\omega
_{\perp }t$, $\widetilde{V}_{tr}\left( r\right) =V_{tr}\left( r\right)
/\hbar \omega _{\perp }=(\widetilde{r}^{2}+\mu \widetilde{r}^{4})/2$,$\
\widetilde{\Omega }=$ $\Omega /\omega _{\perp }$, $\widetilde{L}%
_{z}=L_{z}/\hbar $, and $\widetilde{\psi }_{j}=\psi _{j}a_{0}/\sqrt{N}($ $%
j=1,2)$. Then we obtain the dimensionless coupled Gross-Pitaevskii (GP)
equations in the rotating frame by using a variational method,%
\begin{eqnarray}
i\partial _{t}\psi _{1} &=&\left( -\frac{1}{2}\nabla ^{2}+V_{tr}+\beta
_{11}\left\vert \psi _{1}\right\vert ^{2}+\beta _{12}\left\vert \psi
_{2}\right\vert ^{2}-\Omega L_{z}\right) \psi _{1}  \nonumber \\
&&+\left( \lambda _{x}\partial _{x}+i\lambda _{y}\partial _{y}\right) \psi
_{2},  \label{1-3} \\
i\partial _{t}\psi _{2} &=&\left( -\frac{1}{2}\nabla ^{2}+V_{tr}+\beta
_{22}\left\vert \psi _{2}\right\vert ^{2}+\beta _{12}\left\vert \psi
_{1}\right\vert ^{2}-\Omega L_{z}\right) \psi _{2}  \nonumber \\
&&+\left( -\lambda _{x}\partial _{x}+i\lambda _{y}\partial _{y}\right) \psi
_{1},  \label{1-4}
\end{eqnarray}%
where the tilde is omitted for simplicity. Here $\beta _{jj}$ ( $j=1,2$) and
$\beta _{12}=\beta _{21}$ are the dimensionless intra- and interspecies
coupling strengths. In section 3 of the paper, we will also demonstrate the
ef\/fect of RD-SOC $v_{RD}=-i\hbar (\lambda _{x}\sigma _{x}\partial
_{x}+\lambda _{y}\sigma _{y}\partial _{y})$ \cite{Ozawa,Han} on the ground
state of the system, where the dimensionless dynamic equations with RD-SOC
are expressed by%
\begin{eqnarray}
i\partial _{t}\psi _{1} &=&(-\frac{1}{2}\nabla ^{2}+V_{tr}+\beta
_{11}\left\vert \psi _{1}\right\vert ^{2}+\beta _{12}\left\vert \psi
_{2}\right\vert ^{2}-\Omega L_{z})\psi _{1}  \nonumber \\
&&+(-i\lambda _{x}\partial _{x}-\lambda _{y}\partial _{y})\psi _{2},
\label{1-5} \\
i\partial _{t}\psi _{2} &=&(-\frac{1}{2}\nabla ^{2}+V_{tr}+\beta
_{22}\left\vert \psi _{2}\right\vert ^{2}+\beta _{12}\left\vert \psi
_{1}\right\vert ^{2}-\Omega L_{z})\psi _{2}  \nonumber \\
&&+(-i\lambda _{x}\partial _{x}+\lambda _{y}\partial _{y})\psi _{1}.
\label{1-6}
\end{eqnarray}%
By using the nonlinear Sigma model, we introduce a normalized complex-valued
spinor $\mathbf{\chi }=[\chi _{1},\chi _{2}]^{T}$ with the normalization
condition $|\chi _{1}|^{2}+|\chi _{2}|^{2}=1$. The corresponding component
wave function is $\psi _{j}$ $=\sqrt{\rho }\chi _{j}$ $\left( j=1,2\right) $
and the total density is $\rho =\left\vert \psi _{1}\right\vert
^{2}+\left\vert \psi _{2}\right\vert ^{2}$. The spin density is given by $%
\mathbf{S}=\overline{\mathbf{\chi }}\mathbf{\sigma \chi }$, where $\sigma
=\left( \sigma _{x},\sigma _{y},\sigma _{z}\right) $ are the Pauli matrices.
The components of $\mathbf{S}$ can be written as%
\begin{eqnarray}
S_{x} &=&\frac{1}{\rho }\left( \psi _{1}^{\ast }\psi _{2}+\psi _{2}^{\ast
}\psi _{1}\right) ,  \label{1-7} \\
S_{y} &=&\frac{-i}{\rho }\left( \psi _{1}^{\ast }\psi _{2}-\psi _{2}^{\ast
}\psi _{1}\right) ,  \label{1-8} \\
S_{z} &=&\frac{1}{\rho }\left( \left\vert \psi _{1}\right\vert
^{2}-\left\vert \psi _{2}\right\vert ^{2}\right) ,  \label{1-9}
\end{eqnarray}%
where the norm of the spin is $|\mathbf{S}|=\sqrt{%
S_{x}^{2}+S_{y}^{2}+S_{z}^{2}}=1$. In terms of the above expressions, we
show that the different density distributions of the ground state lead to
different spin density profiles, i.e., different spin textures carrying with
different topological charges. The spatial distribution of topological
structure of the system can be well described by the topological charge
density%
\begin{equation}
q(r)=\frac{1}{4\pi }\mathbf{S\bullet }\left( \frac{\partial \mathbf{S}}{%
\partial x}\times \frac{\partial \mathbf{S}}{\partial y}\right) ,
\label{1-10}
\end{equation}%
and the topological charge is defined as $Q=\int q\left( \mathbf{r}\right)
\mathrm{{d}x{d}y}$. Furthermore, the total topological charge $\left\vert
Q\right\vert $ is conserved if one exchanges $S_{x}$, $S_{y}$, and $S_{z}$.

\section{Results and discussion}

In order to study the ground-state structures of the system and the
corresponding spin textures, we numerically calculate the two-dimensional
(2D) coupled GP Equations (\ref{1-3})-(\ref{1-6}) in terms of the
imaginary-time propagation method \cite{Wen1,WuB,Wen2} based on the
split-step Fourier method \cite{ZhangY}. Recently, phase diagrams for a BEC
with Rashba SOC (R-SOC) in a nonrotating harmonic trap and in a rotating one
have been given in Ref. \cite{Aftalion}. In the present work, we
systematically investigate the combined effects of rotation, SOC and
interatomic interactions on the ground states of the BECs in an anharmonic
trap. In our simulation, we assume the intra- and interspecies interactions
to be repulsive. Without loss of generality, the anharmonicity parameter of
the external trap is chosen as $\mu =1/2$, and the intracomponent
interaction strengths are fixed as $\beta _{11}=\beta _{22}=200$. For
convenience, we introduce a relative interaction strength, $\delta =\beta
_{12}\diagup \beta _{11}$, where the initial state corresponds to an
immiscible state when $\delta >1$ (initially immiscible), while it
corresponds to a miscible state when $\delta <1$ (initially miscible). It is
shown that system can exhibit intriguing properties which are inaccessible
in other systems.

\subsection{Effect of rotation in the absence of SOC}

In Fig. 1 we show the density profiles (the first two rows) and the
corresponding phase profiles (the middle two rows) for the ground states of
spin-$1/2$ BECs in a rotating anharmonic trap, and the last row represents $%
\left\vert \psi _{1}\right\vert ^{2}+\left\vert \psi _{2}\right\vert ^{2}$.
Here $\delta =0.5$, $\lambda _{x}=\lambda _{y}=0$, and the rotation
frequencies in columns (a)-(d) are $\Omega =0.5$, $\Omega =1.65$, $\Omega
=1.7$ and $\Omega =3.0$, respectively. For relatively small rotation
frequency $\Omega =0.5$, a singly quantized vortex occurs spontaneously in
each component (see Fig. 1(a)). The system begins to exhibit partial phase
separation in spite of the two components being mixed initially, which is
caused by the competition among the total interaction energy, the kinetic
energy (especially for the rotational kinetic energy) and the anharmonic
external potential \cite{Wen1}. When $\Omega =1.65$, stable square vortex
lattice forms in each component, where the vortices in the two components
are spatially staggered duo to the repulsive interspecies interaction. As
rotation frequency increases, the number of visible vortices gradually
increases and the square vortex lattice evolves into a triangular vortex
lattice (see Fig. 1(c)). This property is nontrivial as it has not been
reported before. According to previous literature \cite{Kasamatsu2,Mason},
only the structural phase transition from triangular vortex lattice to
square vortex lattice can possibly occur in two-component BECs in a rotating
harmonic trap. Our numerical simulation shows that the trend of the
structural phase transition is irrelevant to the relative interaction
strength $\delta $. The main reason for this difference is the presence of
anharmonic trap which makes the regimes of ultrafast rotation $\Omega >1$
and new phenomena become accessible. With the further increase of rotation
frequency, e.g., $\Omega =3$, we can find that the visible vortices form a
vortex necklace, where the vortices are distributed along a ring. In
particular, as displayed in Fig. 1(d), hidden vortices \cite%
{Wen3,Wen4,Mithun,Price} show up in the central region of the atom cloud and
they form a hidden giant vortex plus a hidden vortex necklace (i.e., an
annular hidden vortex lattice), which is remarkably different from the
topological structures of the BECs in a rotating harmonic potential \cite%
{XuX,Zhou,Radic,Aftalion,Kasamatsu2,Mason}. In the latter case, the central
density hole is a simple giant vortex.

%%%%%%%%%%%%%%%%%%%%%%%%%%%%%%%%%%%%
\begin{figure}[tbh]
\centerline{\includegraphics*[width=\linewidth]{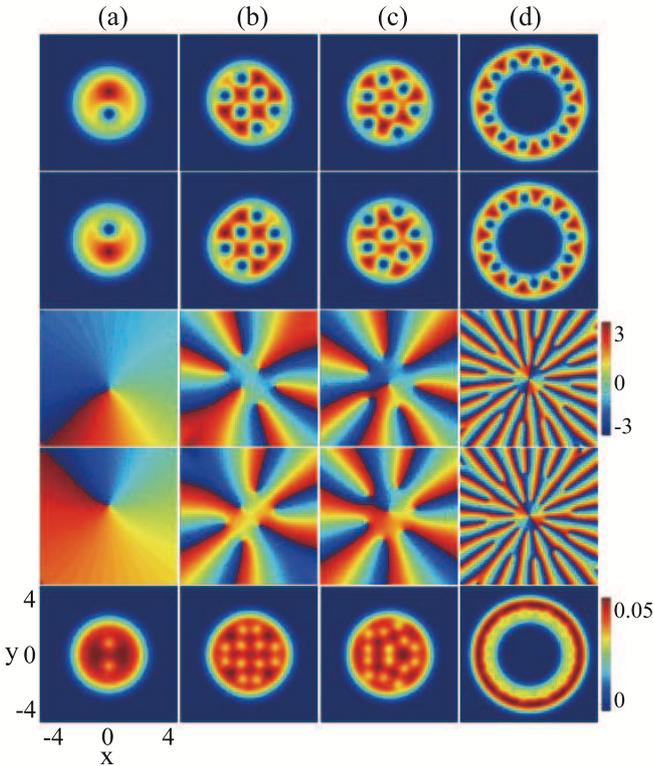}}
%\linespread{0.8}
\caption{(Color online) Ground states of interacting spin-1/2 BECs without
SOC in a rotating anharmonic trap, where $\protect\lambda _{x}=\protect%
\lambda _{y}=0$ and $\protect\delta =0.5$. (a) $\Omega =0.5$, (b) $\Omega
=1.65$, (c) $\Omega =1.7$, and (d) $\Omega =3$. The rows from top to bottom
represent $\left\vert \protect\psi _{1}\right\vert ^{2}$, $\left\vert
\protect\psi _{2}\right\vert ^{2}$, $\arg \protect\psi _{1}$, $\arg \protect%
\psi _{2}$ and $\left\vert \protect\psi _{1}\right\vert ^{2}+$ $\left\vert
\protect\psi _{2}\right\vert ^{2}$, respectively. The horizontal and
vertical coordinates $x$ and $y$ are in units of $a_{0}$.}
\label{figure1}
\end{figure}
%%%%%%%%%%%%%%%%%%%%%%%%%%%%%%%%%%%%

Besides the line-like vortex excitation with respect to the spatial degrees
of freedom of the BECs, the spin degrees of freedom allow for point-like
topological excitation (skyrmion excitation) whose structure and topology
are fixed by simple energy stability. Although skyrmions are already known
in nuclear physics \cite{Skyrme}, quantum Hall systems \cite{Schmeller}, and
liquid crystals \cite{Wright}, generating these topological excitations in
ultracold atomic gases would provide a new opportunity for understanding
their physical properties in much greater detail. In addition, the skyrmion
excitations in the ultracold atomic gases enable a thorough comparison
between the experiments and relevant theories due to the ultrahigh purity
and precise controllability of the cold-atom system. Obviously, the skyrmion
excitations do not have an analogy in a scalar BEC. Shown in Figs. 2(a) and
2(b) are the topological charge density and the spin texture for the
parameters in Fig. 1(b). The local enlargement of the spin texture is given
in Figs. 2 (c) and 2(d). In Figs. 2(e) and 2(f) we display the spin texture
and the local enlargement for the parameters in Fig. 1(d), respectively. The
color of each arrow in the spin textures indicates the magnitude of $S_{z}$.
According to Ref. \cite{LiuC1}, there are seven basic types of skyrmions in
BECs in view of the skyrmion solution with $|S|^{2}=1$: radial-out skyrmion,
radial-in skyrmion, circular skyrmion, hyperbolic skyrmion,
hyperbolic-radial(out) skyrmion, hyperbolic-radial(in) skyrmion and
circular-hyperbolic skyrmion. From Fig. 2(b) we can see that there are a
series of circular-hyperbolic skyrmions (Fig. 2(c)) and
hyperbolic-radial(out) skyrmions (Fig. 2(d)) \cite{LiuC1} which collectively
constitute a composite skyrmion lattice. For the larger rotation frequency $%
\Omega =3$, the spin texture becomes more complex and the skyrmions form two
concentric annular skyrmion lattices as shown in Fig. 2(e), where the
skyrmion configurations include multiple basic types \cite{LiuC1}. The local
circular-hyperbolic skyrmion in Fig. 2(f)) is similar to that in Fig. 2(c)
except for the opposite circular spin current direction. Our calculation
shows that the topological charge of each skyrmion is $\left\vert
Q\right\vert =1$, which is consistent with the definition \cite%
{Skyrme,LiuC1,Zhai}.

In general, Dzyaloshinskii-Moriya interaction plays a key role in the
formation of magnetic skyrmions in chiral magnets, which is induced by the
relativistic spin-orbit coupling \cite{Qaiumzadeh,Ninomiya}. The typical
skyrmions in magnetic materials are Bloch-type skyrmion, N\'{e}el-type
skyrmion, and intermediate-type skyrmion \cite{Kezsmarki,Ahmed}. However,
the skyrmions in trapped BECs are expected to display more novel properties
and richer structures due to the precise controllability of many
experimental parameters of ultracold atoms. Essentially, the skyrmions in
the spin textures of the two-component BECs are associated with the vortices
in the component density distributions, and the particle density must
satisfy the continuity condition as a result of quantum fluid nature of the
BECs. Our simulation shows that the skyrmions in BECs are jointly determined
by multiple parameters such as interatomic interactions, rotation frequency,
SOC, and external potential. In contrast to magnetic materials, SOC is not a
necessary condition for the formation of skyrmions in multi-component BECs.
For rotating two-component BECs without SOC, one can define three new
coupling constants: $c_{0}=(\beta _{11}+\beta _{22}+2\beta _{12})/4$, $%
c_{1}=(\beta _{11}-\beta _{22})/2$, and $c_{2}=(\beta _{11}+\beta
_{22}-2\beta _{12})/4$ \cite{Kasamatsu,Kasamatsu3}. The coefficient $c_{1}$
may be regarded as a longitudinal (pseudo) magnetic field that aligns the
spin along the $z$ axis. The coefficient $c_{2}$ describes the spin-spin
interaction associated with $S_{z}$, where the system is ferromagnetic for $%
c_{2}<0$ and antiferromagnetic for $c_{2}>0$. Thus skyrmions can be
generated in rotating and interacting two-component BECs without SOC. The
detailed discussion can be found in related literature \cite%
{Kasamatsu,Kasamatsu3}. As mentioned above, the skyrmion with fixed unit
topological charge $\left\vert Q\right\vert =1$ can display different
configurations. As pointed out in Ref. \cite{LiuC1}, the
hyperbolic-radial(out) skyrmion, hyperbolic-radial(in) skyrmion, and
circular-hyperbolic skyrmion have two extreme values of $S_{z}$: a minima
and a maximum. The radial-in skyrmion, radial-out skyrmion, circular
skyrmion, and hyperbolic skyrmion can be distinguished by using three
characteristic numbers: polarity $p=sgn\left[ \mathbf{e}_{z}\cdot \mathbf{S}%
(r=0)\right] $, circulation $c=sgn\left\{ \mathbf{e}_{z}\cdot \left[ \mathbf{%
r\times S}(r\neq 0)\right] \right\} $, and divergence $d=sgn\left[ \mathbf{e}%
_{r}\cdot \mathbf{S}(r\neq 0)\right] $ \cite{Wintz}. In the following
sections, one can see that the other parameters, such as the SOC, also
significantly influence the structure of formed skyrmions (see Figs. 2, 6
and 8). As a matter of fact, the components of spin density $S_{x}$, $S_{y}$%
, and $S_{z}$ depend on the component wave functions $\psi _{1,2}$ and the
corresponding complex conjugates $\psi _{1,2}^{\ast }$ of the ground state
of the system as shown in Eqs. (\ref{1-7})-(\ref{1-9}). Therefore the
complex skyrmion structures can be formed in the spin textures for different
combinations of parameters.

%%%%%%%%%%%%%%%%%%%%%%%%%%%%%%%%%%%%
\begin{figure*}[tbh]
\centerline{\includegraphics*[width=0.8\linewidth]{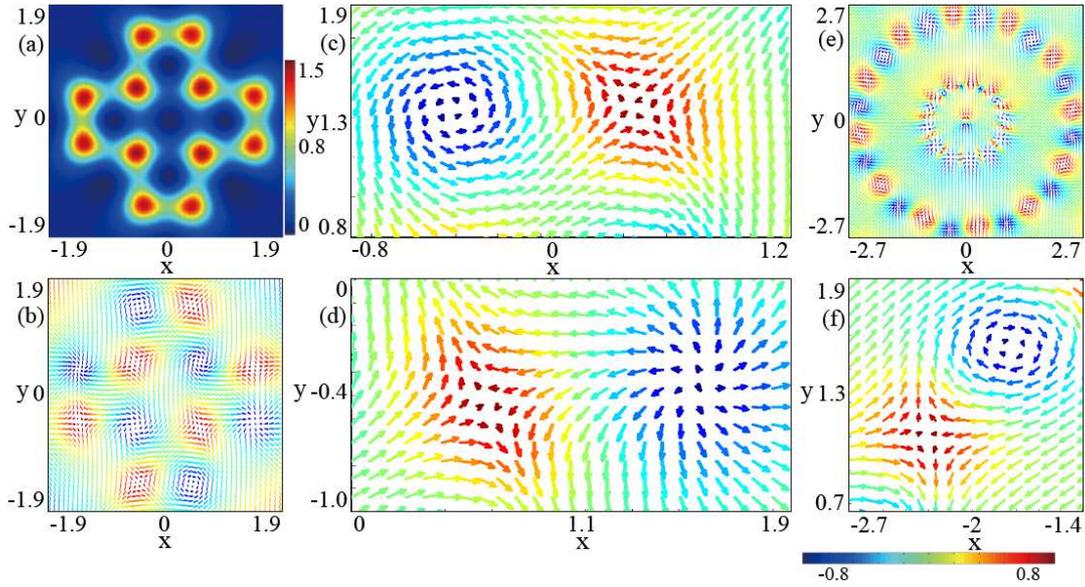}}
\caption{(Color online) Topological charge density and spin textures of
spin-1/2 BECs without SOC in a rotating anharmonic trap, where $\protect%
\lambda _{x}=\protect\lambda _{y}=0$, and $\protect\delta =0.5$. (a)
Topological charge density for $\Omega =1.65$, (b) the corresponding spin
texture, and (c)-(d) local enlargement of the spin texture, where the ground
state is given in Fig. 1(b). (e) Spin texture for $\Omega =3$ and (f) local
enlargement of the spin texture, where the ground state is shown in Fig.
1(d). The color of each arrow indicates the magnitude of $S_{z}$. The
horizontal and vertical coordinates $x$ and $y$ are in units of $a_{0}$.}
\label{figure2}
\end{figure*}
%%%%%%%%%%%%%%%%%%%%%%%%%%%%%%%%%%%%

Note that sometimes the hyperbolic-radial(out) skyrmion,
hyperbolic-radial(in) skyrmion and circular-hyperbolic skyrmion can also be
called a pair of merons (half-skyrmions) which is early proposed in the
study of superfluid Helium with particular cylindrical boundary conditions
\cite{Volovik} and is associated with the so-called Mermin-Ho vortices \cite%
{Mermin}. The topological charges of a meron (half-skyrmion) and an
anti-meron (half-antiskyrmion) are $1/2$ and $-1/2$, respectively. Recent
investigation shows that not all cases of the above three types of skyrmions
can be regarded as meron pairs \cite{LiuC1}. For the meron pair, the
topological charge density is distributed anisotropically along the
polarization direction of the meron pair, and there is a configuration of
vortex-antivortex pair in the relative phase distribution of the two
(pseudo)spin components, where the two vortices are connected by a domain
wall with fixed phase difference of $2\pi $ \cite{Kasamatsu,Kasamatsu3}. In
addition, the spin vector for a meron (half-skyrmion) covers a half of a
unit sphere of spin space while that for a skyrmion sweeps the whole unit
sphere of spin space, which is due to the fact that their topological
charges are $1/2$ and $1$, respectively. In other words, for a certain unit
cell, a skyrmion means that if the spin-density components $S_{x}$ and $S_{y%
\text{ }}$can vary from $-1$ to $1$, then the spin-density component $S_{z%
\text{ }}$also varies from $-1$ to $1$. By comparison, a meron (half
skyrmion) implies that the spin-density component $S_{z}$ just varies from $%
-1$ to $0$ or from $0$ to $1$. In our numerical calculations we identify
skyrmions and merons (half skyrmions) by combining density distribution,
phase distribution, relative phase distribution (when necessary), spin
texture, three spin-density components, topological charge density, local
topological charge in a certain unit cell, and the above features of
skyrmions and merons (half skyrmions).

\subsection{Combined effect of SOC, rotation and interatomic interactions}

\subsubsection{Fixed isotropic D-SOC}

Figure 3 shows the density distributions (the left two columns) and phase
distributions (the right two columns) for the ground states of rotating
anharmonic spin-$1/2$ BECs with relatively small isotropic D-SOC, $\lambda
_{x}=\lambda _{y}=1$, where the odd and even columns denote component $1$
and component $2$, respectively. The rotation frequencies for the initial
component mixing with $\delta =0.5$ in rows (a) and (b) are $\Omega =0.5$
and $\Omega =3$, and those for the component separation with $\delta =2$ in
rows (c)-(f) are $\Omega =0.2$, $\Omega =0.6$, $\Omega =2$ and $\Omega =3$,
respectively. For the initially miscible BECs, when the rotation frequency
is small there is a visible vortex in the center of component 1 and there
are two visible vortices in component 2, where the three vortices
alternatively arrange into a straight line (see Fig. 3(a)). When the
rotation frequency increases to $\Omega =3$, each component generates a
complex topological structure composed of an annular visible vortex lattice
(i.e., a visible vortex necklace) and a large density hole. Here the large
density hole denotes a hidden giant vortex plus a hidden annular vortex
lattice (hidden vortex necklace) rather than a pure giant vortex predicted
in rotating conventional BECs \cite{Fetter2} or rotating Rashba spin-orbit
coupled BECs in a harmonic trap \cite{XuX,Zhou,Radic,Aftalion}. Furthermore,
the density profiles and phase profiles for the two components become
similar except that there is a hidden giant vortex with six circulation
quanta in the center of component 1 while one with seven circulation quanta
in the center of component 2. Although giant vortices are not easy to be
observed in rotating conventional BECs due to their instability \cite%
{Fetter3}, our present results indicate that they can be easily detected in
rotating spin-orbit coupled BECs in the anharmonic trap. Thus the
topological structure of the system is strongly affected by the interplay
among the SOC, rotation frequency, the interatomic interactions, and the
anharmonic trap.

For the initially immiscible BECs, when the rotation frequency is small,
e.g., $\Omega =0.2$ and $\Omega =0.6$, the two components exhibit obvious
phase separation as shown in Figs. 3(c) and 3(d). The topological structures
of the system in Figs. 3(c) and 3(d) are typical Anderson-Toulouse coreless
vortices \cite{Anderson} (some recent literature also called them
half-quantum vortices \cite{Ramachandhran,Zhou}), where the core of the
circulating external component is filled with the other nonrotating
component. With the further increase of rotation frequency, component 1
evolves into a special topological configuration which is composed of an
exterior vortex necklace and a hidden triangular vortex lattice in the
central and very narrow region, while component 2 develops into a
topological structure comprised of an outer annular vortex lattice and an
inner square vortex lattice (see Fig. 3(e)). Here the two components still
keep good phase separation, and the topological defects essentially form
alternative circular vortex sheets. When the rotation frequency increases to
$\Omega =3$, the visible vortices in each component form an annular vortex
lattice, which is similar to the case of initial phase mixing. The region of
the large density hole is occupied by a central hidden giant vortex (a
doubly quantized vortex for component 1 and a triply quantized vortex for
component 2) and a hidden vortex necklace. Therefore one can conclude that
when the rotation is relatively small the topological defects of the ground
state are jointly determined by the interatomic interactions, SOC, rotation
frequency, and anharmonic confinement. When $\Omega $ is large enough,
however, rotation plays a crucial role in the topological defect formation
of the system.

%%%%%%%%%%%%%%%%%%%%%%%%%%%%%%%%%%%%
\begin{figure}[tbh]
\centerline{\includegraphics*[width=\linewidth]{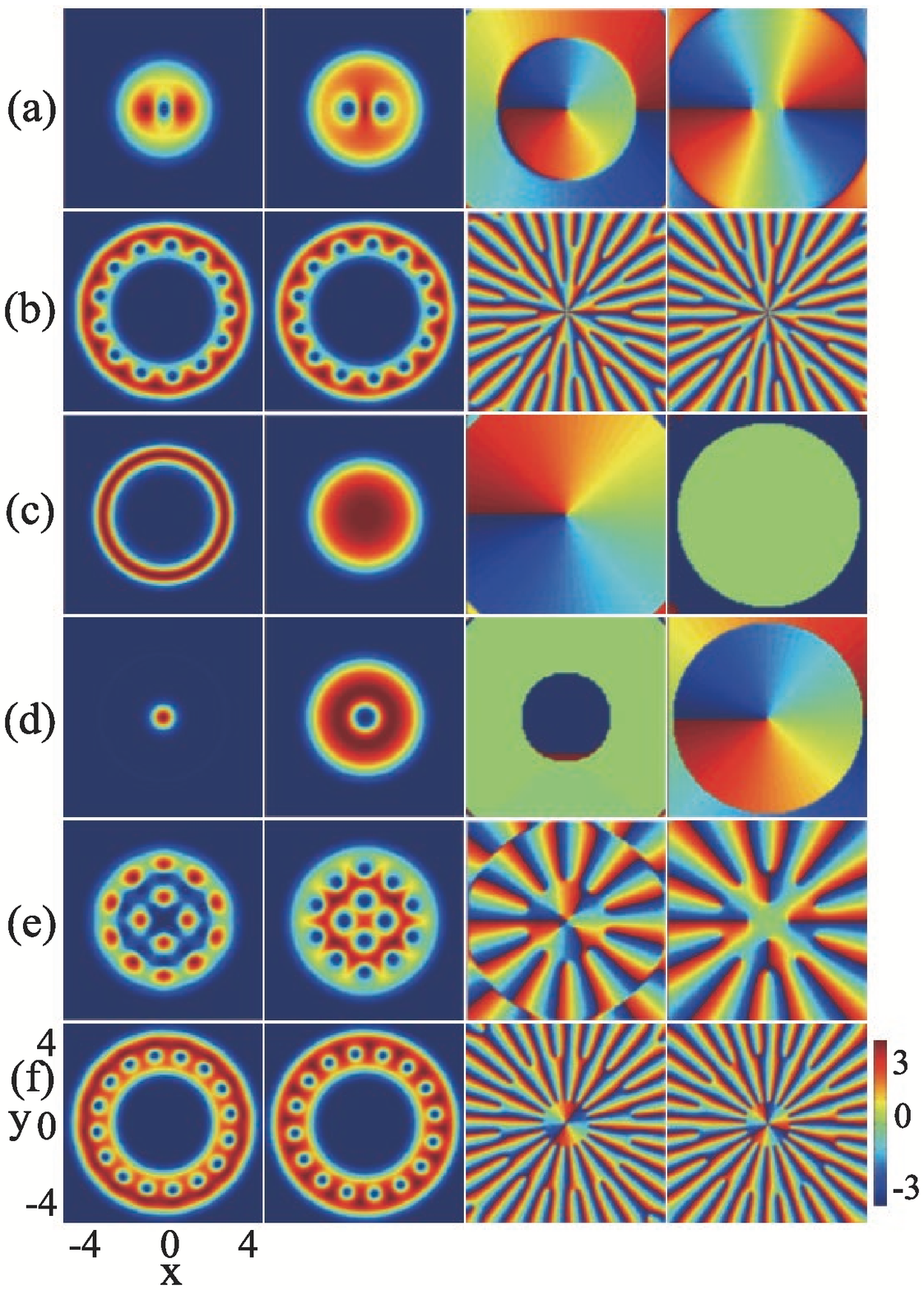}}
\caption{(Color online) Ground-state density distributions and phase
distributions for rotating anharmonic spin-$1/2$ BECs with $\protect\lambda %
_{x}=\protect\lambda _{y}=1$. (a) $\protect\delta =0.5$, $\Omega =0.5$, (b) $%
\protect\delta =0.5$, $\Omega =3$, (c) $\protect\delta =2$, $\Omega =0.2$,
(d) $\protect\delta =2$, $\Omega =0.6$, (e) $\protect\delta =2$, $\Omega =2$%
, and (f) $\protect\delta =2$, $\Omega =3$. The columns from left to right
denote $\left\vert \protect\psi _{1}\right\vert ^{2}$, $\left\vert \protect%
\psi _{2}\right\vert ^{2}$, arg$\protect\psi _{1}$, and arg$\protect\psi %
_{2} $, respectively. The horizontal and vertical coordinates $x$ and $y$
are in units of $a_{0}$.}
\label{figure3}
\end{figure}
%%%%%%%%%%%%%%%%%%%%%%%%%%%%%%%%%%%%

Displayed in Figs. 4(a) and 4(c) are the topological charge densities for
the parameters in Figs. 3(a) and 3(d), respectively. The corresponding spin
textures are shown in Figs. 4(b) and 4(d). The spin texture in Fig. 4(b)
represents a special topological structure with topological charge
approaching $Q=3/2$ which is comprised of a radial-out meron in the center
and two symmetric hyperbolic merons in the two sides. The exotic spin defect
may be called a meron cluster consisting of three merons. Shown in Fig. 4(d)
is a typical hyperbolic skyrmion with unit topological charge $Q=1$ whose
ground-state structure generally corresponds to an Anderson--Toulouse
coreless vortex.

%%%%%%%%%%%%%%%%%%%%%%%%%%%%%%%%%%%%
\begin{figure}[tbh]
\centerline{\includegraphics*[width=\linewidth]{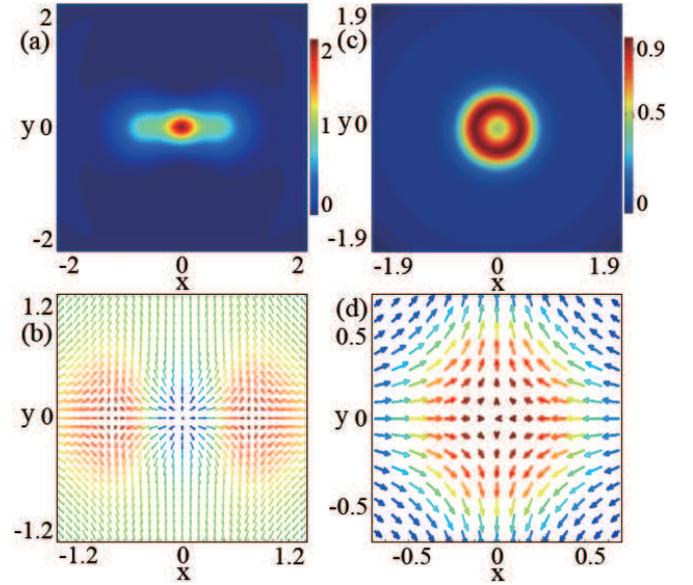}}
\caption{(Color online) Topological charge densities (a), (c)and spin
textures (b), (d) for the ground states of interacting spin-$1/2$ BECs in a
rotating anharmonic trap, where $\protect\lambda _{x}=\protect\lambda _{y}=1$%
. (a)-(b) $\protect\delta =0.5$, $\Omega =0.5$, (c)-(d) $\protect\delta =2$,
$\Omega =0.6$. The horizontal and vertical coordinates $x$ and $y$ are in
units of $a_{0}$.}
\label{Figure4}
\end{figure}
%%%%%%%%%%%%%%%%%%%%%%%%%%%%%%%%%%%%

\subsubsection{Fixed rotation frequency}

Our calculations show that in the absence of rotation the large SOC
strengths favor vortex chains for initially mixed BECs and stripe phases as
well as heliciform-stripe phases for initially demixed BECs. For relatively
small rotation frequency $\Omega <1$, the ground states of the system are
somewhat similar to those of spin-orbit coupled BECs in a rotating harmonic
trap \cite{XuX,Zhou,Aftalion}. For the sake of brevity, we will not repeat
these results in this work. Here we mainly focus on the case of large
rotation frequency, i.e. $\Omega >1$. Figure 5 shows the density
distributions and phase distributions of the system with fixed rotation
frequency $\Omega =1.2$ above the radial trapping frequency, which is
usually not accessible in conventional rotating BECs confined in a harmonic
trap \cite{XuX,Zhou,Radic,Aftalion,Fetter,LiuC1,Kasamatsu2}. The columns
from left to right represent $\left\vert \psi _{1}\right\vert ^{2}$, $%
\left\vert \psi _{2}\right\vert ^{2}$, arg$\psi _{1}$, and arg$\psi _{2}$,
respectively. Here the strengths of the 2D isotropic D-SOC for the initial
component mixing with $\delta =0.5$ in rows (a)-(c) and those for the
initial component separation with $\delta =2$ in rows (d) and (e) are $%
\lambda _{x}=\lambda _{y}=$ $0.8$ and $\lambda _{x}=\lambda _{y}=$ $25$,
respectively. In the case of initial component mixing and weak D-SOC, four
visible vortices arrange into a square vortex lattice in component 1, while
in component 2 the five visible vortices constitute a criss-cross vortex
string, which reduces effectively the energy of system to the minimum (see
Fig. 5(a)). Essentially, the visible vortices in the system form a coreless
vortex lattice as a whole because the vortices are alternatively distributed
in the two components and the remainder of the two components keep mixing,
which means that there is no topological defect in the total density
distribution. With the increase of D-SOC strength, e.g., $\lambda
_{x}=\lambda _{y}=$ $2$, there exists a topological phase transition for the
vortex structure of the system. From Fig. 5(b), the vortex number in each
component increases evidently because the stronger SOC means the larger
orbital angular momentum input into the system. In addition, the vortices in
each component form an annular vortex lattice plus a central vortex (a
singly-quantized central vortex for component 1 while a doubly-quantized
central vortex for component 2). As a result, the visible vortices compose a
nucleated vortex lattice which is remarkably different from Fig. 5(a)
because there are evident vortex defects in the total density distribution
of the system.

For a strong D-SOC with $\lambda _{x}=\lambda _{y}=$ $25$, there is an
almost full overlap of the density distributions and the phase distributions
between the two components. This character occurs for not only the case of
initial phase mixing but also the case of initial phase separation (see
Figs. 5(c) and 5(e)). Here the visible vortices constitute multilayer
ringlike structures. Our simulation shows that the region of the large
density hole is occupied by a central hidden giant vortex and several hidden
vortex necklaces (see Figs. 5(c) and 5(e)), which is quite different from
the conventional prediction results in rotating spin-orbit coupled BECs in a
harmonic trap \cite{XuX,Zhou,Radic,Aftalion,Fetter}. For the latter case,
the large density hole in the density distribution at large rotation
frequency corresponds to a pure giant vortex. It is shown that the present
anharmonic system can exhibit rich and complex topological configuration due
to the interplay among interatomic interactions, rotation, SOC, and
anharmonic confinement.

In the case of initial component separation and weak D-SOC, the two
component densities exhibit obvious phase separation as shown in Fig. 5(d),
where the visible vortices or hidden vortices in the two components tend to
form annular structures layer by layer. With the further increase of D-SOC,
e.g., $\lambda _{x}=\lambda _{y}=$ $25$, more vortices generate in the cloud
and form a rather complex topological structure comprised of three-layer
visible vortex necklaces and several hidden vortex necklaces as well as a
central hidden giant vortex (see Fig. 5(e)), which is similar to that in
Fig. 5(c). Physically, the strong D-SOC or the rapid rotation will result in
large kinetic energy. Here the kinetic energy acts against the interspecies
interaction. The latter is responsible for component demixing while the
former tends to expand the BECs and hence favors component mixing. In the
mean time, the anharmonic trapping potential tends to trap the BECs more
tightly and thus also sustains component mixing. Thus component separation
can be suppressed by the kinetic energy and external potential in some
conditions even if the relation $\beta _{11}\beta _{22}<\beta _{12}^{2}$ is
satisfied, as we can see in Figs. 1, 2 and 5.\ In addition, we can find that
when relevant parameters such as interatomic interactions and rotation
frequency are fixed the D-SOC can be used to control the topological
structure of the rotating anharmonic BECs.

%%%%%%%%%%%%%%%%%%%%%%%%%%%%%%%%%%%%
\begin{figure}[tbh]
%\begin{figure*}
\centerline{\includegraphics*[width=\linewidth]{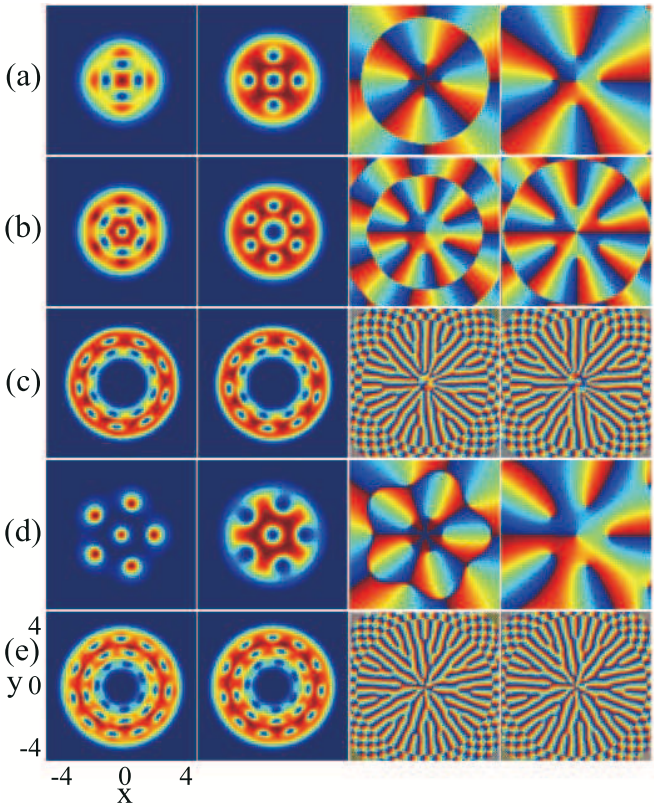}}
\caption{(Color online) Ground states of rotating interacting spin-$1/2$
BECs with D-SOC in an anharmonic trap, where $\Omega =1.2$. (a) $\protect%
\delta =0.5$, $\protect\lambda _{x}=\protect\lambda _{y}=0.8$, (b) $\protect%
\delta =0.5$, $\protect\lambda _{x}=\protect\lambda _{y}=2$, (c) $\protect%
\delta =0.5$, $\protect\lambda _{x}=\protect\lambda _{y}=25$, (d) $\protect%
\delta =2$, $\protect\lambda _{x}=\protect\lambda _{y}=0.8$, and (e) $%
\protect\delta =2$, $\protect\lambda _{x}=\protect\lambda _{y}=25$. The
columns from left to right denote $\left\vert \protect\psi _{1}\right\vert
^{2}$, $\left\vert \protect\psi _{2}\right\vert ^{2}$, arg$\protect\psi _{1}$
and arg$\protect\psi _{2}$, respectively. The horizontal and vertical
coordinates $x$ and $y$ are in units of $a_{0}$.}
\label{Figure5}
\end{figure}
%%%%%%%%%%%%%%%%%%%%%%%%%%%%%%%%%%%%

The topological charge density and the spin texture for the parameters in
Fig. 5(a) are demonstrated in Fig. 6. From Fig. 6(a), the topological charge
density exhibits a petal-like structure composed of four evident criss-cross
petals and an inconspicuous torus in the trap center. Figure 6(b) is the
corresponding spin texture, where the local amplifications are displayed in
Figs. 6(c) and 6(d). The four skyrmions and one half-skyrmion (meron) in
Fig. 6(b) constitute a novel skyrmion lattice in which the vertical two spin
defects are hyperbolic-radial(in) skyrmions, the horizontal two spin defects
are hyperbolic-radial(out) skyrmions, and the central spin defect is a
hyperbolic half-skyrmion (meron) \cite{Skyrme,LiuC1,Kasamatsu}. Our
nonlinear stability analysis shows that the vortex structures, skyrmion
structures and topological properties of the system are rather stable when
the SOC strength or\ the rotation frequency is not too large. In addition,
we find that as the SOC strength or rotation frequency increases the
topological charge of the system is increases first and then decreases. For
very strong D-SOC or very fast rotation, however, the skyrmion lattice
configuration in spin textures will be destroyed. Physically, when the SOC
strength (with fixed rotation frequency) or the rotation frequency (with
fixed SOC) increases, more angular momentum contributes to the system and
leads to generation of more vortices, regardless of the initial state of the
system being mixed or separated. In this case most vortices in component 1
and those in component 2 are separated from each in space due to the
repulsive interspecies interaction, which means that the total topological
charge of the system continues to increase. However, when the SOC strength
or the rotation frequency become very large the system favors an almost full
overlap between the densities of the two components because higher angular
momentum states become dominant as seen in Figs. 1, 3 and 5, which indicates
the topological density $q(\mathbf{r})$ decreases remarkably due to $%
S_{z}\rightarrow 0$. According to Equation (\ref{1-10}), it is easy to prove
that the topological charge density is strictly zero for any planar spin
texture. Thus the total topological charge decreases dramatically for the
case of very large SOC strength or rotation frequency.

%%%%%%%%%%%%%%%%%%%%%%%%%%%%%%%%%%%%
\begin{figure}[tbh]
%\begin{figure*}
\centerline{\includegraphics*[width=\linewidth]{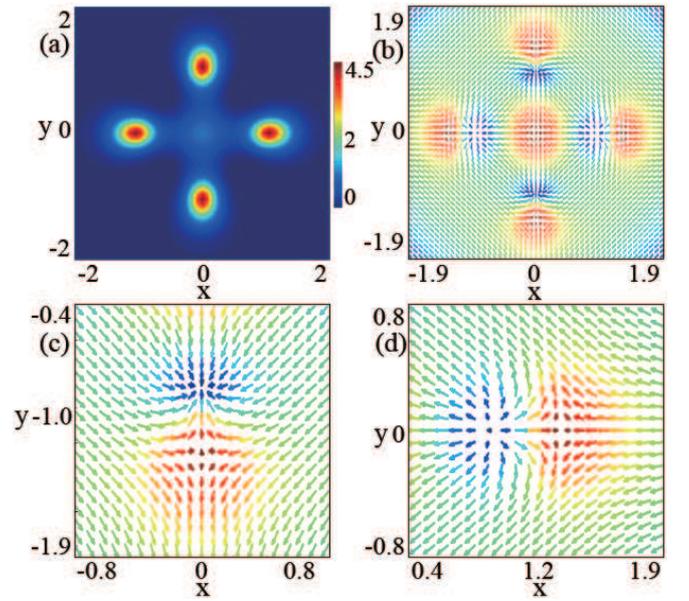}}
\caption{(Color online) Topological charge density (a) and spin textures
(b)-(d), where $\protect\delta =0.5$, $\protect\lambda _{x}=\protect\lambda %
_{y}=0.8$, and $\Omega =1.2$. The corresponding ground state is shown in
Fig. 5(a). (c) and (d) are the local amplifications of the spin texture. The
horizontal and vertical coordinates $x$ and $y$ are in units of $a_{0}$.}
\label{figure6}
\end{figure}
%%%%%%%%%%%%%%%%%%%%%%%%%%%%%%%%%%%%

\subsection{Effects of anisotropic D-SOC and RD-SOC}

Next, we consider the ground-state properties of rotating anharmonic BECs
with anisotropic D-SOC and those with anisotropic RD-SOC. The D-SOC and
RD-SOC can be realized within current experimental techniques \cite%
{Lin,Cheuk,Meng,ZWu,HuangL,Kolkowitz,LiJ,Qu,Ozawa,Han}, and we will also
discuss the difference between the D-SOC effect and the RD-SOC effect. In
the isotropic case of $\lambda _{x}=\lambda _{y}$, the RD-SOC belongs to the
R-SOC type, so we may only consider the anisotropic case. Figure 7 shows the
density distributions (the left columns of (a)-(f)) and phase distributions
(the right columns of (a)-(f)) of component 1 for the ground states of
rotating anharmonic spin-$1/2$ BECs with anisotropic D-SOC or RD-SOC. Note
that Figs. 7(a), 7(b), 7(d) and 7(e) denote the cases of D-SOC and the
others represent the cases of RD-SOC.

%%%%%%%%%%%%%%%%%%%%%%%%%%%%%%%%%%%%
\begin{figure}[tbh]
%\begin{figure*}
\centerline{\includegraphics*[width=\linewidth]{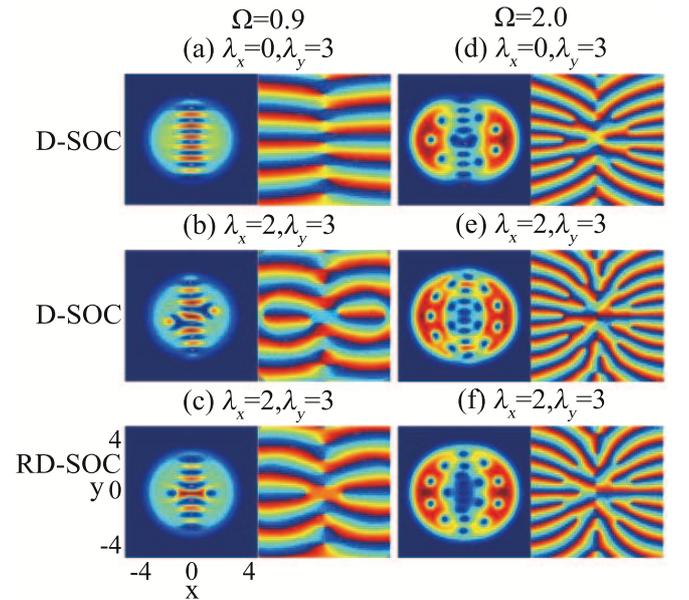}}
\caption{(Color online) Density distributions $\left\vert \protect\psi %
_{1}\right\vert ^{2}$ (the left columns of (a)-(f)) and phase distributions $%
\arg \protect\psi _{1}$ (the right columns of (a)-(f)) for the ground states
of rotating anharmonic spin-$1/2$ BECs with anisotropic D-SOC and those with
anisotropic RD-SOC, where $\protect\delta =2.0$. (a), (b), (d) and (e) are
the cases of anisotropic D-SOC, while (c) and (f) are the cases of
anisotropic RD-SOC. (a) $\Omega =0.9$, $\protect\lambda _{x}=0$, $\protect%
\lambda _{y}=3$, (b)-(c) $\Omega =0.9$, $\protect\lambda _{x}=2$, $\protect%
\lambda _{y}=3$, (d) $\Omega =2.0$, $\protect\lambda _{x}=0$, $\protect%
\lambda _{y}=3$, and (e)-(f) $\Omega =2.0$, $\protect\lambda _{x}=2$, $%
\protect\lambda _{y}=3$. The horizontal and vertical coordinates x and y are
in units of $a_{0}$.}
\label{figure7}
\end{figure}
%%%%%%%%%%%%%%%%%%%%%%%%%%%%%%%%%%%%

In the case of $\Omega =0.9$, for 1D D-SOC there is an obvious vortex chain
along the $x=0$ axis because of the 1D spin-orbit interaction along the $y$
direction as shown in Fig. 7(a). The density distribution exhibits well
symmetry with respect to the $x$ $=0$ axis and the $y=0$ axis. Component 2
has similar density distribution and phase distribution except that the
vortices in the two components are staggered in space, which is not shown
here for the sake of simplicity. When $\lambda _{y}$ is enhanced, only the
number of vortices along the $x=0$ axis increases without changing the
symmetry of the density distribution. For 2D anisotropic D-SOC, there is no
density symmetry with respect to the $x$ $=0$ axis or the $y=0$ axis (see
Fig. 7(b)). By comparison, for the case of anisotropic RD-SOC, the system
keeps well the symmetry with respect to $x=0$ axis and $y=0$ axis,
especially for the latter case and for the exterior region in the density
distribution (see Figs. 7(c) and 7(f)). Similar properties also exist for
large rotation frequency, e.g., $\Omega =2$. In Fig. 7(d), vortices are
distributed not only along the $y$ direction but also along the $x$
direction, where the symmetry with respect to $x=0$ axis is broken while the
symmetry with respect to $y=0$ axis are basically unchanged. The physical
mechanism is that for large rotation frequency the $x$-direction vortex
chain caused by the combined effect of 1D D-SOC and rotation can only carry
finite angular momentum and the residual angular momentum is inevitably
carried by the transverse vortices beside the $x=0$ axis. For 2D anisotropic
D-SOC with $\lambda _{x}=2$ and $\lambda _{y}=3$, more vortices are
generated in the system and there is a distorted vortex chain along the $y$
direction, but both the density symmetry concerning the $x=0$ axis and that
concerning the $y=0$ axis are destroyed, where the outer vortices tend to
form an annular vortex array (Fig. 7(e)). The main reason is that the
kinetic energy plays a key role in the ground-state structure of the system
for large rotation frequency. When the SOC is Rashba-Dresselhaus type (see
Fig. 7(f)), the density distribution restores the excellent symmetry
concerning the $y=0$ axis, which is similar to Fig. 7(c).

%%%%%%%%%%%%%%%%%%%%%%%%%%%%%%%%%%%%
\begin{figure}[tbh]
%\begin{figure*}
\centerline{\includegraphics*[width=\linewidth]{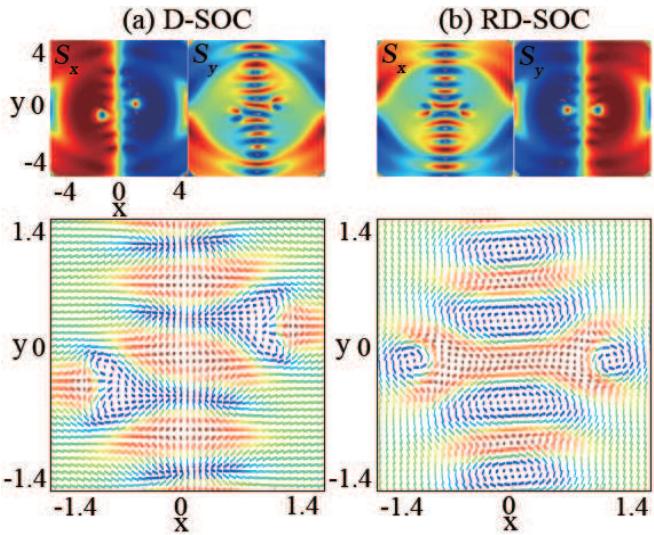}}
\caption{(Color online) The components $S_{x}$ and $S_{y}$ (the top row) of
spin density vector and the spin textures (the bottom row), where $\protect%
\delta =2.0$, $\Omega =0.9$, $\protect\lambda _{x}=2$, and $\protect\lambda %
_{y}=3$. (a) and (b) denote the case of anisotropic D-SOC and that of
anisotropic RD-SOC, respectively. The horizontal and vertical coordinates x
and y are in units of $a_{0}$.}
\label{figure8}
\end{figure}
%%%%%%%%%%%%%%%%%%%%%%%%%%%%%%%%%%%%

The $S_{x}\ $and $S_{y}$ components of spin density vector and the
corresponding spin textures for the parameters in Figs. 7(b) and 7(c) are
displayed in Fig. 8. In the pseudo-spin representation, the blue region
denotes spin-down and the red region denotes spin-up. In the case of D-SOC, $%
S_{x}$ and $S_{y}$ obey neither even nor odd parity distribution along the $%
x $ direction or the $y$ direction (Fig. 8(a)). By comparison, in the case
of RD-SOC, $S_{x}$ satisfies the even parity distribution along the $x$
direction and the odd parity distribution along the $y$ direction, while it
is just the reverse for $S_{y}$ (Fig. 8(b)). At the same time, the spin
component $S_{x}$ ($S_{y}$) in Fig. 8(a) (Fig. 8(b)) forms two remarkable
spin domains, and the boundary between the two spin domain develops into a
domain wall with $\left\vert S_{x}\right\vert \neq 1$ ($\left\vert
S_{y}\right\vert \neq 1$). Generally, the spin domain wall for nonrotating
two-component BECs is a classical N\'{e}el wall, where the spin flips only
along the vertical direction of the wall. However, our numerical simulation
of the spin texture shows that in the region of spin domain wall the spin
flips not only along the $x$ direction (the vertical direction of domain
wall) but also along the $y$ direction (the domain-wall direction), which
indicates that the observed spin domain wall in the present system is a new
type of domain wall. Here the anisotropic SOC leads to the creation of more
interesting and complicated skyrmion structures. This feature is present
evidently for both the cases of anisotropic D-SOC and anisotropic RD-SOC.
For the case of D-SOC, there is an obvious radial(in) skyrmion string along
the $x=0$ axis with two asymmetric hyperbolic skyrmions standing on either
side of the axis. By comparison, a circular skyrmion string is distributed
along the $x=0$ axis with two symmetric circular skyrmions lying on either
side of the axis. Note that the ground-state structures and spin textures
for the case of anisotropic SOC have not been discussed in the previous
investigations on rotating spin-orbit coupled quantum gases \cite%
{XuX,Zhou,Radic,Aftalion,Fetter}. We expect that the interesting topological
defects and spin textures found in the present work allow to be demonstrated
in the future experiments.

\section{Conclusion}

We have studied the ground-state structures of interacting two-component
BECs with D-SOC or RD-SOC in a rotating anharmonic trap. The effects of
rotation, SOC, interatomic interaction, and trap anharmonicity on the
ground-state structure of this system are analyzed and discussed
systematically. In the absence of SOC, with the increase of rotation
frequency the system experiences structural phase transitions from singly
quantized vortex state to square vortex lattice, square vortex lattice to
triangular vortex lattice, and triangular vortex lattice to ringlike vortex
lattice. In the presence of isotropic D-SOC, the system supports vortex
pair, Anderson--Toulouse coreless vortices, circular vortex sheets and
combined complex topological structures comprised of a visible vortex
necklace and a central hidden giant vortex plus a hidden vortex necklace,
where the latter does not depend on the interaction parameters. In addition,
large D-SOC yields a visible multi-layer vortex necklace and a central
hidden giant vortex as well as several hidden vortex necklaces. Furthermore,
it is shown that the system sustains single basic skyrmion, meron cluster,
skyrmion string and various complex skyrmion lattices including criss-cross
skyrmion lattices and multi-ring skyrmion lattices. In the limit of large
SOC or rotation frequency, the skyrmion configurations in spin textures will
be destroyed due to the enhanced overlap between the two component densities
and thus the distinct reduction of topological charge density. Moreover, the
effects of anisotropic D-SOC and RD-SOC on the topological structures of the
system are discussed. Compared with the case of anisotropic D-SOC,\ the
system for the case of anisotropic RD-SOC maintains well the symmetry with
respect to $x=0$ and $y=0$ axes, especially for the latter case and for the
outer region in the density distribution. New domain wall and skyrmion
structures are revealed in the cases of anisotropic SOC. The exotic
topological defects and spin textures can be tested and observed in the
future experiments, and thus the work presents fantastic perspective for
topological excitations in cold atom physics and condensed matter physics.

\begin{acknowledgment}

L.W. thanks Chuanwei Zhang, Hui Zhai, Yongping Zhang, Zhi-Fang Xu,
Xiang-Fa Zhou,and Malcolm Jardine for helpful discussions, and acknowledges the
research group of Professor W. Vincent Liu at the University of
Pittsburgh, where part of the work was carried out. In addition, we give special thanks to
Professor Ran Cheng at the University of California, Riverside for his productive discussions
and insightful suggestions. This work is
supported by the National Natural Science Foundation of China (Grant
No. 11475144), the Natural Science Foundation of Hebei Province of
China (Grant No. A2015203037), and Research Foundation of Yanshan
University (Grant No. B846).

\end{acknowledgment}

\end{document}